\documentclass[conference]{IEEEtran}
\usepackage{graphicx, array, amssymb, psfrag, graphics, setspace, stfloats}
\usepackage{amsfonts}
\usepackage[caption=false,font=normalsize,labelfont=sf,textfont=sf]{subfig}
\usepackage[toc,abbreviations,nonumberlist,nogroupskip,nopostdot]{glossaries-extra}
\glssetcategoryattribute{general}{glossdesc}{firstuc}
\glssetcategoryattribute{abbreviation}{glossdesc}{title}
\usepackage{mathtools}
\usepackage{multirow}
\usepackage{amsmath}
\usepackage{bibentry}

\DeclarePairedDelimiter\floor{\lfloor}{\rfloor}

\DeclareMathOperator*{\argmin}{arg\,min}

\newcommand{\m}[1]{\mathbf{#1}}
\newcommand{\E}[1]{\mathbb{E} \left[{#1}\right]}
\newcommand{\Prob}[1]{\text{Pr} \left({#1}\right)}

\newcommand{\myabs}[1]{\left\lvert#1\right\rvert}

\def\scalingFig{0.27}

\newglossaryentry{computer}
{
name=computer,
description={A programmable machine that receives input data,
  stores and manipulates the data, and provides
  formatted output}
}

\newglossary*{nomenclature}{Nomenclature}

\newabbreviation{ISI}{ISI}					{inter-symbol interference}
\newabbreviation{DAC}{DAC}					{digital-to-analog converter}
\newabbreviation{ADC}{ADC}					{analog-to-digital converter}
\newabbreviation{DSP}{DSP}					{digital signal processing}
\newabbreviation{TX}{TX}					{transmitter}
\newabbreviation{RX}{RX}					{receiver}
\newabbreviation{PSK}{PSK}					{phase shift keying}
\newabbreviation{QAM}{QAM}					{quadrature-amplitude modulation}
\newabbreviation{FEC}{FEC}					{forward error correction}
\newabbreviation{SOP}{SOP}					{state-of-polarization}
\newabbreviation{FF}{FF}					{feed-forward}
\newabbreviation{FFEQ}{FFEQ}				{feed-forward equalizer}
\newabbreviation{BER}{BER}					{bit error rate}
\newabbreviation{SNR}{SNR}					{signal-to-noise ratio}
\newabbreviation{RSNR}{RSNR}				{required SNR}
\newabbreviation{SNDR}{SNDR}				{signal-to-noise-and-distortion ratio}
\newabbreviation{SFDR}{SFDR}				{spurious free dynamic range}
\newabbreviation{RC}{RC}					{raised cosine}
\newabbreviation{RRC}{RRC}					{root raised cosine}
\newabbreviation{ENOB}{ENOB}				{effective number of bits}
\newabbreviation{GD}{GD}					{group delay}
\newabbreviation{CMD}{CMD}					{chromatic dispersion}
\newabbreviation{PMD}{PMD}					{polarization mode dispersion}
\newabbreviation{PDL}{PDL}					{polarization dependent loss}
\newabbreviation{ASE}{ASE}					{amplified spontaneous emission}
\newabbreviation{LMS}{LMS}					{least mean squares}
\newabbreviation{APA}{APA}					{affine projection algorithm}
\newabbreviation{NLMS}{NLMS}				{normalized LMS}
\newabbreviation{MMSE}{MMSE}				{minimum mean square error}
\newabbreviation{CMA}{CMA}					{Constant modulus algorithm}
\newabbreviation{RLS}{RLS}					{recursive least squares}
\newabbreviation{LS}{LS}					{least squares}
\newabbreviation{LO}{LO}					{local-oscillator}
\newabbreviation{CR}{CR}					{carrier-recovery}
\newabbreviation{ASIC}{ASIC}				{application-specific integrated circuits}
\newabbreviation{FIR}{FIR}					{finite impulse response}
\newabbreviation{DD-LMS}{DD-LMS}			{decision-directed least mean squares}
\newabbreviation{DD}{DD}					{decision-directed}
\newabbreviation{CS-DAC}{CS-DAC}			{current-steering DAC}
\newabbreviation{LSB}{LSB}					{least-significant bit}
\newabbreviation{MSB}{MSB}					{most-significant bit}
\newabbreviation{DNL}{DNL}					{differential non-linearity}
\newabbreviation{INL}{INL}					{integral non-linearity}
\newabbreviation{DGD}{DGD}					{differential group delay}
\newabbreviation{FFT}{FFT}					{fast-Fourier transform}
\newabbreviation{IFFT}{IFFT}				{inverse fast-Fourier transform}
\newabbreviation{DFT}{DFT}					{discrete Fourier transform}
\newabbreviation{IDFT}{IDFT}				{inverse discrete Fourier transform}
\newabbreviation{FT}{FT}					{Fourier transform}
\newabbreviation{MSE}{MSE}					{mean square error}
\newabbreviation{HD}{HD}					{hard decision}
\newabbreviation{SD}{SD}					{soft decision}
\newabbreviation{LDPC}{LDPC}				{low density parity check}
\newabbreviation{CW}{CW}					{continuous wave}
\newabbreviation{PBC}{PBC}					{polarization beam combiner}
\newabbreviation{MIMO}{MIMO}				{multiple-input and multiple-output}
\newabbreviation{OPGW}{OPGW}				{optical ground wire}
\newabbreviation{ZF}{ZF}					{zero-forcing}
\newabbreviation{CAZAC}{CAZAC}				{Constant amplitude zero auto-correlation}
\newabbreviation{CFO}{CFO}					{carrier frequency offset}
\newabbreviation{MA}{MA}					{moving average}
\newabbreviation{DE}{DE}					{Differential evolution}
\newabbreviation{SA}{SA}					{simulated annealing}
\newabbreviation{DEM}{DEM}					{Dynamic element matching}
\newabbreviation{LUT}{LUT}					{lookup table}
\newabbreviation{DP}{DP}					{dynamic programming}
\newabbreviation{DPC}{DPC}					{digital pre-compensation}
\newabbreviation{NN}{NN}					{neural network}
\newabbreviation{MLSE}{MLSE}				{maximum–likelihood sequence estimation}
\newabbreviation{LE}{LE}					{linear equalizer}
\newabbreviation{DFE}{DFE}					{decision–feedback equalizer}
\newabbreviation{THP}{THP}					{Tomlinson-Harashima precoding}
\newabbreviation{HW}{HW}					{hardware}
\newabbreviation{PS}{PS}					{pilot sequence}
\newabbreviation{SW-LS}{SW-LS}				{sliding window least squares}
\newabbreviation{RD-Kalman}{RD-Kalman}      {radius-directed Kalman}	
\newabbreviation{RMS}{RMS}                  {root mean square}
\newabbreviation{SQNR}{SQNR}                {signal-to-quantization-noise ratio}
\newabbreviation{PDF}{PDF}                  {probability density function}
\newabbreviation{CDF}{CDF}                  {cumulative distribution function}
\newabbreviation{AWGN}{AWGN}			    {additive white Gaussian noise}

\newglossaryentry{dingledorf}
{
type=nomenclature,
name=dingledorf,
description={A person of supposed average intelligence who makes incredibly brainless misjudgments}
}

\newglossary*{symbols}{List of Symbols}
\newglossaryentry{rvec}
{
name={$\mathbf{v}$},
sort={label},
type=symbols,
description={Random vector: a location in n-dimensional Cartesian space, where each dimensional component is determined by a random process}
}
\newglossaryentry{C}
{
name={\ensuremath{\mathcal{C}}},
sort={label},
type=symbols,
description={Configuration space}
}
\newglossaryentry{C-space}
{
name={\ensuremath{\mathcal{C}\text{-space}}},
sort={label},
type=symbols,
description={Configuration space}
}
\newglossaryentry{C-free}
{
name={\ensuremath{\mathcal{C}^{\text{free}}}},
sort={label},
type=symbols,
description={Configuration space free of constraints/obstacles}
}
\newglossaryentry{P}
{
name={\ensuremath{\mathcal{P}}},
sort={label},
type=symbols,
description={Path planning space}
}
\newglossaryentry{P-constraint}
{
name={\ensuremath{\mathcal{P}^{\text{constraint}}}},
sort={label},
type=symbols,
description={Path planning space corresponding to constraints (including obstacles)}
}
\newglossaryentry{P-free}
{
name={\ensuremath{\mathcal{P}^{\text{free}}}},
sort={label},
type=symbols,
description={$\equiv \mathcal{P} \setminus \mathcal{P}^{\text{constraint}}$ Path planning space free of constraints/obstacles}
}
\newglossaryentry{p-goal}
{
name={\ensuremath{\textbf{p}^{\text{goal}}}},
sort={label},
type=symbols,
description={Destination point}
}
\newglossaryentry{p-start}
{
name={\ensuremath{\textbf{p}^{\text{start}}}},
sort={label},
type=symbols,
description={Starting point}
}
\newglossaryentry{W}
{
name={\ensuremath{\mathcal{W}}},
sort={label},
type=symbols,
description={Workspace}
}

\graphicspath{{Fig/}} 
\begin{document}

\title{Timing-Error Optimized Architecture for Current-Steering DACs}

\author{Ramin~Babaee, \textit{Student Member}, \textit{IEEE}, Shahab Oveis Gharan, and Martin Bouchard, \textit{Senior Member}, \textit{IEEE}\\
	\thanks{}}

 \maketitle

\begin{abstract}
We propose a novel \gls*{DAC} weighting architecture that statistically minimizes the distortion caused by random timing mismatches among current sources. To decode the DAC input codewords into corresponding DAC switches, we present three algorithms with varying computational complexities. We perform high-level Matlab simulations to illustrate the dynamic performance improvement over the segmented structure. 
\end{abstract}


\section{Introduction}
\IEEEPARstart{W}{ith} the rapid growth of high-speed communication systems, high resolution, and high data rate DACs are a necessity. Current-steering DAC is a prominent architecture that enables the use of DACs in high-speed applications \cite{9863991, toumazou_1993}. 

The \gls*{DAC} linear performance is influenced by both static and dynamic distortion sources. The current cell mismatches introduce amplitude errors that affect the static performance. However, the timings skew among current switches impacts the frequency-dependent dynamic performance. 
As the data rate increases, the signal degradation arising from timing mismatches becomes the leading impairment \cite{7932085}. 


The redundancy offered by the segmented structure provides an excellent opportunity for utilizing two classes of mitigation techniques. \gls*{DEM} randomizes the selection of unary cells. This algorithm improves the \gls*{SFDR} performance by converting the frequency-dependent distortion into white noise. However, the total \gls*{SNDR} remains unchanged \cite{911479,476173,5420027,RTSC,913021,9485113}. Calibration schemes address the mitigation of mismatch errors by impairment estimation followed by a correction. The studies in \cite{4362086,34100,1528680} focus on amplitude mismatch mapping, while \cite{1692532, 5751593} investigates a mapping technique for reducing timing errors. 

In our paper \cite{AE_paper}, we introduced the concept of weighting optimization for minimizing statistical amplitude errors. This study focuses on statistically random timing errors where the induced glitch errors are proportional to the weight of the current source. 
We propose a general weighting scheme that decides on the current levels of different DAC switches in order to minimize DAC dynamic distortion sources. 

The rest of the paper is organized as follows. In Section \ref{sec_te:te_glitch_analysis}, a detailed analysis of random DAC timing errors is presented. Section \ref{sec_te:te_architrecture_design} discusses the proposed DAC weighting scheme. Representation selection methods that decode the input codewords to the corresponding DAC switches are presented in Section \ref{sec_te:te_rep_sel}. 
Simulation results are provided in Section \ref{sec_te:te_simulaion_results} and followed by conclusions in Section \ref{sec_te:te_conclusion}.

\textbf{Notation:} Vectors and scalars are represented by bold letters and non-bold italic letters, respectively. $\m{B}_i$ represents the $i$-th element of vector $\m{B}$. $(\cdot)^\text{T}$ indicates transpose of a vector/matrix. 
Symbol $\E{\cdot}$ is used for statistical expectation operation.
The cardinality of set $\mathcal{R}$ is also denoted by $|\mathcal{R}|$. 

\section{Timing Glitch Analysis}
\label{sec_te:te_glitch_analysis}

\begin{figure}[!tb]
\centering
\footnotesize
\psfrag{Switch ON transient}{Switching on transient}
\psfrag{Switch OFF transient}{Switching off transient}
\psfrag{time}{Time [s]}
\psfrag{T}{$T$}
\psfrag{0}{$0$}
\psfrag{1}{$1$}
\psfrag{Bi}{$\m{B}_i$}
\psfrag{Output current}{\hspace{-6ex} Normalized output current} 
\includegraphics[scale=\scalingFig]{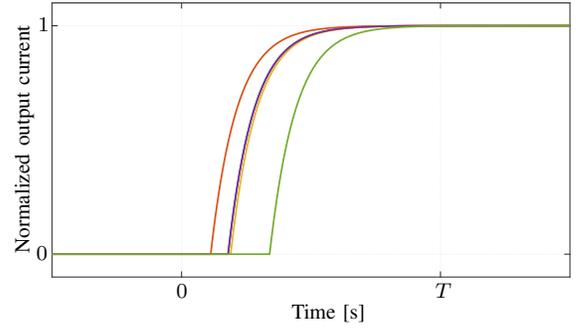}
\caption{Timing error in settling behavior of \gls*{DAC} switches.}
\label{fig:te_time}
\end{figure}

Let $N$ and $L$ denote the number of input bits and current switches of a DAC, respectively. The weights of current sources are represented by an $L$-dimensional basis vector $\m{B}$, and the binary representation of a given integer input $x$ in the range $[0 \quad 2^N-1]$ is expressed by an $L$-dimensional binary vector $\m{W}(x)$. Note that $\m{B}_i = 2^i$ and $\m{B}_i =1$ for a binary DAC and a unary DAC, respectively. Consequently, the current output of the DAC is formulated as
\begin{equation}
x = \m{W}^\text{T}(x) \m{B}.
\end{equation}

A transition from input $x$ to $y$ may cause a change in the state of each switch of the DAC, which can be formulated as
\begin{equation}
 \m{c}(x,y) = \m{W}(y) - \m{W}(x) ,
\end{equation}
where $\m{c}(x,y)$ is a vector with $L$ elements and $\m{c}_i(x,y) \in \{+1,0,-1\}, \; 0 \le i\le L-1 $. 
A timing skew vector $\m{\tau}$ of size $L$ is defined to represent the transition time of the switches. Ideally, all the values should be $0$. But, due to component mismatch, the values are distributed between $-\frac{1}{2}$ to $\frac{1}{2}$, where they are normalized by the sampling period $T$. Statistically, $\m{\tau}_i$ values are assumed to be random, independent, and identically distributed, and have a Gaussian distribution with a mean of zero and a variance of $\sigma^2_\m{\tau}$. Fig. \ref{fig:te_time} illustrates timing errors among \gls*{DAC} switches using an exponential transient response. Although an exponential function is more realistic, for the purpose of the analysis in this section, we assume the desired transient response of \gls*{DAC} switches is an ideal step function $u(t)$ with transition time at $0$, i.e.,
\begin{equation}
u(t) =
\begin{cases}
0, \hspace{10ex} & t \le 0, \\
1, \hspace{10ex} & t> 0. \\
\end{cases}
\end{equation}
It is worth mentioning that a step function transition model may not be an accurate model of the hardware, but it allows for an efficient formulation of the glitch error and optimizing the DAC architecture. The simulation results show that this simplification can still provide a very significant improvement.

Assuming switching on and off transients have the same response, the \gls*{DAC} output can be written as
\begin{equation}
z(t) = x + \sum_{i=0}^{L-1} \m{c}_i(x,y) \m{B}_i u(t-\tau_i),   \hspace{4ex} -\frac{T}{2} < t \le \frac{T}{2} .
\end{equation}
Given that the ideal transition time is $0$, one can derive the transient error at the output of the \gls*{DAC} as
\begin{equation}
e_g (t) = \sum_{i=0}^{L-1} \m{c}_i(x,y) \m{B}_i \Pi (\frac{t-\tau_i/2}{\myabs{\m{\tau}_i}}) \text{sgn}(\tau_i), 
\label{eq_te:te_err1}
\end{equation}
where $\Pi(t)$ is the rectangular function defined as
\begin{equation}
\Pi(t) =
\begin{cases}
0, \hspace{10ex}            & |t|>\frac{1}{2},  \\
\frac{1}{2}, \hspace{10ex}  & |t|=\frac{1}{2},  \\
1, \hspace{10ex}            & |t|<\frac{1}{2},  \\
\end{cases}
\end{equation}
and $\text{sgn}(t)$ is the sign function. Note that for notation simplicity, we drop the argument $(x,y)$ from $\m{c}_i$ terms. Without loss of generality, we assume $\tau_i$ values are sorted such that $\m{\tau}_0 \le \m{\tau}_1 \le \dots \le \tau_{L-1}$. Eq. (\ref{eq_te:te_err1}) can be then expanded as
\begin{equation}
e_g (t) = 
    \begin{cases}
     -\m{c}_0 \m{B}_0, \hspace{50ex}                & \m{\tau}_0 \le t < \m{\tau}_1,         \\
     -\m{c}_0 \m{B}_0 - \m{c}_1 \m{B}_1, \quad      & \m{\tau}_1 \le t < \tau_2,             \\
     \hdots                                         & \hdots                                 \\
     -\sum_{i=0}^{m} \m{c}_i \m{B}_i, \quad         & \tau_m \le t < 0,                      \\
     \sum_{i=m}^{L-2} \m{c}_i \m{B}_i, \quad        & 0 \le t < \tau_{m+1},                  \\
     \hdots                                         & \hdots                                 \\
     \m{c}_{L-2} \m{B}_{L-2}, \quad                 & \tau_{L-2} \le t < \tau_{L-1},         \\
    \end{cases} 
\end{equation}
where $\tau_m$ is the largest negative number. The \gls*{MSE} associated with the transition from $x$ to $y$ is then defined as
\begin{align}
C(x,y) & = \frac{1}{T} \E{ \int_{-\frac{T}{2}}^{\frac{T}{2}} \myabs{e_g(t)}^2 dt } \nonumber \\
& = \frac{1}{T} \mathbb{E} \Bigg[ \sum_{n=0}^{m-1} (\sum_{i=0}^{n} \m{c}_i \m{B}_i)^2 (\tau_{n+1}-\tau_n)  \nonumber \\
& \hspace{3ex} - (\sum_{i=0}^{m} \m{c}_i \m{B}_i)^2 \tau_m  + (\sum_{i=m}^{L-2} \m{c}_i \m{B}_i)^2 \tau_{m+1} \nonumber \\
& \hspace{3ex}  + \sum_{n=m+1}^{L-2} (\sum_{i=n}^{L-2} \m{c}_i \m{B}_i)^2 (\tau_{n+1}-\tau_n) \Bigg].
\end{align}
Since the timing errors are independent of the current weights and representation vectors, the metric $C(x,y)$ can be simplified to
\begin{align}
C(x,y) & = \frac{2}{T} \sum_{n=0}^{\floor{L/2}-1} \E{ (\sum_{i=0}^{n} \m{c}_i \m{B}_i)^2 } \E{\tau_{n+1}-\tau_n} \nonumber \\
& \hspace{1ex} - \frac{2}{T}  \E{ (\sum_{i=0}^{\floor{L/2}} \m{c}_i \m{B}_i)^2 } \E{\tau_{\floor{L/2}}} \nonumber \\
       & = 2 \sum_{n=0}^{\floor{L/2}} \E{ (\sum_{i=0}^{n} \m{c}_i \m{B}_i)^2 } \mu_n,
\label{eq_te:te_tot_cost1}
\end{align}
where 
\begin{equation}
\mu_n = 
\begin{cases}
    \frac{1}{T} \E{\tau_{n+1}-\tau_n},      \quad & \text{if } n<\floor{L/2},       \\
    \frac{1}{T} \E{\tau_{\floor{L/2}}},     \quad & \text{if } n=\floor{L/2}.       \\
\end{cases}
\end{equation}
Note that $\mu_n$ only depends on 2 parameters $L$ and $\sigma^2_\m{\tau}$ and it scales linearly with $\sigma_\m{\tau}$. The expression inside the summation can be calculated as
\begin{align}
\E{ (\sum_{i=0}^{n} \m{c}_i \m{B}_i)^2 } & =  \sum_{i=0}^{n} \E{ (\m{c}_i \m{B}_i)^2 } + \sum_{i=0}^{n} \sum_{\substack{j=0 \\ j \neq i } }^{n} \E{ \m{c}_i \m{c}_j \m{B}_i \m{B}_j } \nonumber \\
 & = \frac{n+1}{L} \sum_{i=0}^{L-1} (\m{c}_i \m{B}_i)^2 \nonumber \\
 & \hspace{1ex} + \frac{n(n+1)}{L(L-1)} \sum_{i=0}^{L-1} \sum_{\substack{j=0 \\ j \neq i } }^{L-1} \m{c}_i \m{c}_j \m{B}_i \m{B}_j.
\end{align}
By re-writing Eq. (\ref{eq_te:te_tot_cost1}), upon the above result, we obtain
\begin{align}
C(x,y) & = 2 \sum_{n=0}^{\floor{L/2}} (n+1) (D + n S) \mu_n,
\label{eq_te:te_metrc_x_to_y}
\end{align}
where $D$ and $S$ are defined as
\begin{align}
D & = \frac{1}{L} \sum_{i=0}^{L-1} (\m{c}_i \m{B}_i)^2, \\
S & = \frac{1}{L(L-1)} \sum_{i=0}^{L-1} \sum_{\substack{j=0 \\ j \neq i } }^{L-1} \m{c}_i \m{c}_j \m{B}_i \m{B}_j.
\end{align}


Having obtained the MSE associated with a transition from $x$ to $y$, we are now able to derive the total cost function. If there is only one representation for every input $x$ such that $\m{W}^\text{T}(x) \m{B} = x$, then the basis is complete, e.g. a binary DAC. In this case, the total cost can be calculated as
\begin{equation}
C_\text{total} = \sum_{x=0}^{2^N-1} \sum_{y=0}^{2^N-1} \Prob{x,y} C(x,y), 
\label{eq_te:te_metric_single_rep}
\end{equation}
where $\Prob{x,y}$ is the transition probability from input $x$ to input $y$. 

\section{Architecture Design}
\label{sec_te:te_architrecture_design}

For an over-complete basis, there may be multiple representations for an input value. Let $\mathcal{R}(y)$ denote the set of all possible representations for input $y$, i.e.,
\begin{equation}
\mathcal{R}(y) = \left\{\m{W} | \m{W}^\text{T} \m{B} = y\right\}.
\end{equation}
In the case of an over-complete basis, we have multiple choices for the representation of $y$ among the set $\mathcal{R}(y)$. Here, we propose selecting the representation $\m{W}(y) \in \mathcal{R}(y)$ which minimizes the timing glitch expected distortion power $C(x,y)$. Assuming we make such an optimum selection, the metric $C_\text{total}$ can be calculated as
\begin{align}
C_\text{total} & = \sum_{x=0}^{2^N-1} \sum_{y=0}^{2^N-1} \Prob{x,y} \nonumber \\
& \sum_{\m{W}(x) \in \mathcal{R}(x)} \Prob{\m{W}(x) | x} \min_{\m{W}(y) \in \mathcal{R}(y)} C(\m{W}(x),\m{W}(y)),
\label{eq_te:te_cost}
\end{align}
where $\Prob{\m{W}(x)|x}$ is the probability distribution of representations of $x$. Note that we have changed the notation $C(x,y)$ to $C(\m{W}(x),\m{W}(y))$ to emphasize that the cost associated with a transition from $x$ to $y$ depends on their corresponding representation vectors. Also note that Eq. (\ref{eq_te:te_cost}) is a generalized form of Eq. (\ref{eq_te:te_metric_single_rep}) where for a complete basis, the representation sets $\mathcal{R}(x)$ and $\mathcal{R}(y)$ have a cardinality of $1$ and thus, $\Prob{\m{W}(x)|x}=1$. 

The DAC architecture design can be expressed as finding the DAC current switch levels $\m{B}_0, \m{B}_1, \dots, \m{B}_{L-1}$ such that the distortion power due to timing glitch is minimized. Mathematically saying, given the DAC input bit resolution $N$ and number of DAC switches $L$, we aim to find the over-complete basis $\m{B}$ which minimizes the distortion function, specifically,
\begin{equation}
\m{B}_\text{opt} = \argmin_{\m{B}} C_\text{total} .
\label{eq_te:te_op}
\end{equation}

Note that the optimization expression (\ref{eq_te:te_op}) is a non-linear, non-convex, discrete problem. The dimension of search space is $L$ and each element of vector $\m{B}$ may have an integer value in the range $[1 \hspace{1ex} 2^N]$. Therefore, the maximum size of the search space is $2^{NL}$, which grows exponentially with both $N$ and $L$. Therefore, an exhaustive search is not feasible except for small values of $L$ and $N$.

Using the \gls*{SA} algorithm \cite[Chapter 7]{optimization_book}, the optimized basis for an $8$-bit \gls*{DAC} for a few values of $L$ is computed and is presented in Table \ref{tab:te_basis}. 
 
\begin{table}[!tbh]
\footnotesize
\centering
\begin{tabular}{||l||*{14}{p{0.2em}}||}
\hline
\textbf{Basis Length} & \multicolumn{14}{l||}{\textbf{Optimized Basis}} \\ 
\hline
9  &   1&   2&   4&   8&  16&  31&  43&  69&  81&    &    &    &    &    \\
10 &   1&   2&   4&   8&  16&  21&  31&  39&  62&  71&    &    &    &    \\
11 &   1&   2&   4&   8&  13&  18&  26&  30&  38&  54&  61&    &    &    \\
12 &   1&   2&   4&   8&  11&  16&  20&  25&  27&  35&  48&  58&    &    \\
13 &   1&   2&   4&   8&  14&  16&  18&  21&  25&  27&  30&  33&  56&    \\
14 &   1&   2&   4&   8&  12&  13&  14&  15&  18&  21&  26&  31&  35&  55\\
\hline
\end{tabular}
\vspace{2ex}
\caption{Timing error optimized basis vectors for an $8$-bit DAC.}
\label{tab:te_basis}
\end{table}

\section{Representation Selection}
\label{sec_te:te_rep_sel}
In the previous section, we derived the optimum choice of DAC architecture current sources, i.e., $\m{B}_\text{opt}$. However, given an over-complete basis $\m{B}$, there are multiple projections from the set $\mathcal{R}(x)$ which can represent DAC input value $x$. In this section, we consider three different schemes for choosing the best representation for DAC input, assuming the basis $\m{B}$ is already derived. The methods differ in terms of their computational complexity versus their achieved performance. 

\textit{1) Viterbi (optimal)}: Given an input sequence $x[0]$, $x[1]$, ..., $x[M-1]$, the optimal representations can be chosen such that the statistical error power in Eq. (\ref{eq_te:te_cost}) averaged over transitions of the sequence is minimized. 
The optimization problem is formulated as
\begin{equation}
\argmin_{\m{W}(x[0]),...,\m{W}(x[M-1])} \sum_{m=1}^{M-1} C(\m{W}(x[m-1]),\m{W}(x[m])).
\end{equation}

Each node $(y, m)$ is connected to a node $(y', m+1) \in \mathcal{R}(x[m+1])$ with a directed edge $e$ and an associated cost $c(e)$ to such edge which is defined as $c(e) = C(\m{W}(y),\m{W}(y'))$. The cost of a path $(e_1, e_2,...,,e_M)$ is defined as $c(e_1)+c(e_2)+...+c(e_M)$. It can be verified that the optimum value for the above optimization problem can be derived by finding the minimum cost path from a node at $(y, 0)$ to a node at $(y', M-1)$. Such an optimization problem is a well-known Dynamic programming problem that can be solved using the Viterbi algorithm \cite{18626}. The Viterbi algorithm runs sequentially through the Trellis graph: For iteration $m$, it finds the minimum cost path from nodes at position $0$ to any of the nodes $(y, m)$, assuming the minimum cost paths for nodes $(y', m-1)$ is already derived in the previous iteration. Computational complexity of Viterbi algorithm is equal to $\mathcal{O}(M\E{|\mathcal{R}(x)|^2})$, which is equivalent to $\mathcal{O}(\E{|\mathcal{R}(x)|^2})$ per each DAC sample.



\textit{2) Sequential greedy}: Given the significant computational complexity of the Viterbi approach, one can use a greedy approach to sequentially find the best representation for $x[m]$ which minimizes the transition cost from $x[m-1]$ to $x[m]$, assuming the representation for $x[m-1]$ is already known. More precisely, we have
\begin{equation}
\argmin_{\m{W}(x[m])} C(\m{W}(x[m-1]),\m{W}(x[m])).
\end{equation}
Note that this approach produces a sub-optimum result. Furthermore, computational complexity of this algorithm is $\mathcal{O}(\E{|\mathcal{R}(x)|})$ per sample. Hence, it has a lower computational complexity than the Viterbi algorithm. However, similar to the Viterbi method, this approach requires samples to be decoded sequentially which is not ideal for a highly parallel \gls*{DSP} architecture.

\textit{3) Memoryless representation}: This approach chooses a unique representation $\m{W}(x)$ for each DAC input codeword irrespective of its previous and next sample values, namely,
\begin{equation}
\argmin_{\m{W}(x[m])}  \mathbb{E}_z\big[C(\m{W}(x[m]), \m{W}(z))\big]+\mathbb{E}_z\big[C(\m{W}(z),\m{W}(x[m]))\big] 
\end{equation}
Therefore, there is no dependency between samples, and each DAC input can be processed independently. This approach is ideal for parallel implementation in the hardware. In this case, all codeword mappings can be computed offline and the mapping of the input signal to the current weights is performed by a \gls*{LUT} in hardware. 

\section{Simulation Results}
\label{sec_te:te_simulaion_results}
In this section, simulation results are presented to evaluate the performance of the proposed \gls*{DAC} architecture and mapping scheme. 
The simulations are performed with a high-level model of an $8$-bit \gls*{DAC} implemented in Matlab. The only impairment simulated is the random timing errors among DAC switches. The switch's transient response is modeled as an exponential function. 

\begin{figure}[!tb]
\centering
\footnotesize
\psfrag{Hybrid, best_next}{Segmented DAC}
\psfrag{Optimized, single_rep_opt}{Opt. memoryless}
\psfrag{Optimized, best_next}{Opt. sequential greedy}
\psfrag{Optimized, viterbi}{Opt. Viterbi}
\psfrag{Basis length}{Basis length $L$}
\psfrag{Normalized cost (Simulation)}{\hspace{3ex} Normalized metric}
\includegraphics[scale=\scalingFig]{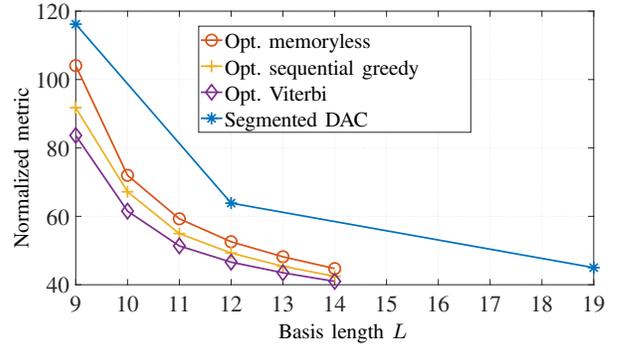}
\caption{Normalized glitch metric as a function of basis length $L$ for the segmented DAC and the three statistical selection algorithms for the optimized architecture.}
\label{fig:te_metric}
\end{figure}

Fig. \ref{fig:te_metric} compares the metric derived in Eq. (\ref{eq_te:te_cost}) for the proposed optimized architecture with the segmented topology. The $x$-axis represents the number of \gls*{DAC} switches, or alternatively basis length $L$, and the $y$-axis is the metric normalized by a fully thermometer-coded \gls*{DAC} metric, i.e., metric is normalized to $1$ for fully thermometer-coded DAC. For the segmented architecture, three configurations are plotted, $2$T+$6$B, $3$T+$5$B, and $4$T+$4$B, where the first number represents the number of thermometer-coded \gls*{MSB}s and the second number denotes the number of binary bits. The number of \gls*{DAC} switches for the segmented case is $9$, $12$, and $19$, respectively. In general, as the number of switches increases, the glitch performance improves significantly. As evident in the figure, the optimized architecture achieves a similar glitch metric with only $13$ switches, when used with the Viterbi algorithm. The greedy algorithm and the memoryless method with 14 switches outperform the $4$T+$4$B architecture. 

\begin{figure}[!tb]
\centering
\footnotesize
\psfrag{Hybrid20, single_rep}{Segmented DAC}
\psfrag{Optimized20, single_rep_opt}{Opt. memoryless}
\psfrag{Optimized20, best_next}{Opt. sequential greedy}
\psfrag{Optimized20, viterbi}{Opt. Viterbi}
\psfrag{SDR [dB]}{SNDR [dB]}
\psfrag{basis_len}{Basis length $L$}
\psfrag{on_off_offset}{\hspace{-10ex} Normalized timing error $(\tau_\text{on/off}/T)$}
\includegraphics[scale=\scalingFig]{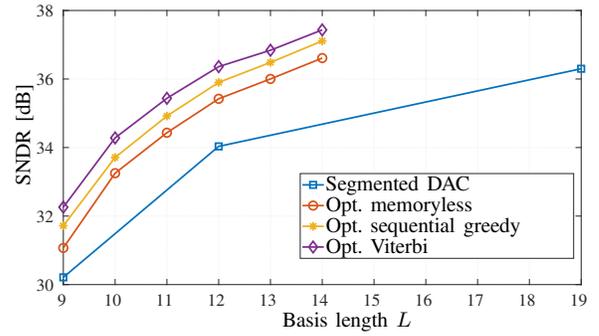}
\caption{Mean glitch SNDR comparison of segmented DAC and the proposed optimized architecture versus basis length $L$.}
\label{fig:te_mean_sndr}
\end{figure}

\begin{figure}[!tb]
\centering
\footnotesize
\psfrag{Hybrid20, single_rep}{Segmented DAC}
\psfrag{Optimized20, single_rep_opt}{Opt. memoryless}
\psfrag{Optimized20, best_next}{Opt. sequential greedy}
\psfrag{Optimized20, viterbi}{Opt. Viterbi}
\psfrag{SDR [dB], Yield = 95}{\hspace{2ex} SNDR [dB]}
\psfrag{basis_len}{Basis length $L$}
\psfrag{on_off_offset}{\hspace{-10ex} Normalized timing error $(\tau_\text{on/off}/T)$}
\includegraphics[scale=\scalingFig]{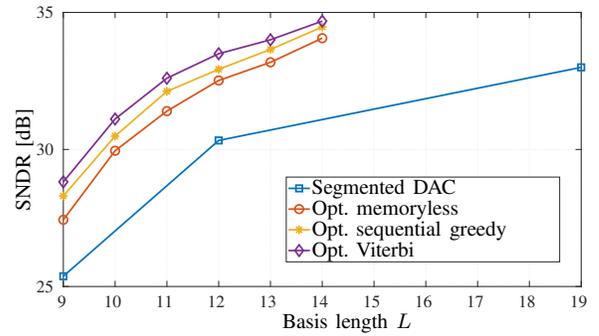}
\caption{Glitch SNDR 95th percentile comparison of segmented DAC and the proposed optimized architecture versus basis length $L$.}
\label{fig:te_95_sndr}
\end{figure}

Using a high-level \gls*{DAC} model, the \gls*{SNDR} achieved by each of the architectures for $\sigma_\tau = 0.03$ is illustrated in Fig. \ref{fig:te_mean_sndr}. The \gls*{SNDR} is defined as the ratio of input signal power to the glitch error power at the output of the DAC and does not include quantization noise. Note that the choice of $\sigma_\tau$ does not impact the relative performance of these two schemes since the glitch power linearly scales with $\sigma^2_\tau$. The results are obtained by averaging error power over 10,000 realizations of random timing skews. The figure clearly demonstrates the advantage of the optimized architectures over the segmented configuration. As an example, in order to achieve $36$dB SNDR, segmented DAC requires 19 switches, while the optimized DAC basis with the Viterbi coding requires $12$ switches. Figure \ref{fig:te_95_sndr} depicts SNDR achieved for $95\%$ yield target for different DAC architectures versus the basis length.

\section{CONCLUSION}
\label{sec_te:te_conclusion}
We studied the glitches caused by random timing errors of \gls*{DAC} switches. We formulated the \gls*{MSE} and derived an optimization problem that was used to find architectures that attain significant improvement over segmented DACs. We then studied the input signal decoding for the proposed DAC architecture and proposed three decoding algorithms with different levels of computational complexity versus attained performance. The proposed memoryless mapping technique offers a simple implementation through \gls*{LUT}, which is suitable for high-speed applications. Although the paper focused on SNDR improvement, one can use the same framework for deriving superior architectures using other metrics such as \gls*{SFDR}. In the next step of this research, we will conduct transistor-level simulations to verify the potential benefit of the proposed DAC architecture.

\bibliography{IEEEabrv,main}

\begin{thebibliography}{10}
\providecommand{\url}[1]{#1}
\csname url@samestyle\endcsname
\providecommand{\newblock}{\relax}
\providecommand{\bibinfo}[2]{#2}
\providecommand{\BIBentrySTDinterwordspacing}{\spaceskip=0pt\relax}
\providecommand{\BIBentryALTinterwordstretchfactor}{4}
\providecommand{\BIBentryALTinterwordspacing}{\spaceskip=\fontdimen2\font plus
\BIBentryALTinterwordstretchfactor\fontdimen3\font minus \fontdimen4\font\relax}
\providecommand{\BIBforeignlanguage}[2]{{%
\expandafter\ifx\csname l@#1\endcsname\relax
\typeout{** WARNING: IEEEtran.bst: No hyphenation pattern has been}%
\typeout{** loaded for the language `#1'. Using the pattern for}%
\typeout{** the default language instead.}%
\else
\language=\csname l@#1\endcsname
\fi
#2}}
\providecommand{\BIBdecl}{\relax}
\BIBdecl

\bibitem{9863991}
G.~Manganaro, ``An introduction to high data rate current-steering {Nyquist} {DACs}: Fasten your seat belts,'' \emph{{IEEE} Solid-State Circuits Mag.}, vol.~14, no.~3, pp. 24--40, 2022.

\bibitem{toumazou_1993}
C.~Toumazou, J.~B. Hughes, and N.~C. Battersby, Eds., \emph{Switched-Currents: An Analogue Technique for Digital Technology}.\hskip 1em plus 0.5em minus 0.4em\relax London, U.K: Peter Peregrinus Ltd., 1993.

\bibitem{7932085}
S.~M. McDonnell, V.~J. Patel, L.~Duncan, B.~Dupaix, and W.~Khalil, ``Compensation and calibration techniques for current-steering {DACs},'' \emph{{IEEE} Circuits Syst. Mag.}, vol.~17, no.~2, pp. 4--26, 2017.

\bibitem{911479}
M.~Vesterbacka, M.~Rudberg, J.~Wikner, and N.~Andersson, ``Dynamic element matching in {D/A} converters with restricted scrambling,'' in \emph{Proc. 7th IEEE Int. Conf. Electron., Circuits and Syst. (ICECS)}, vol.~1, 2000, pp. 36--39.

\bibitem{476173}
R.~Baird and T.~Fiez, ``Linearity enhancement of multibit {$\Delta \Sigma$} {A/D} and {D/A} converters using data weighted averaging,'' \emph{{IEEE} Trans. Circuits Syst. {II}}, vol.~42, no.~12, pp. 753--762, 1995.

\bibitem{5420027}
I.~Galton, ``Why dynamic-element-matching {DACs} work,'' \emph{{IEEE} Trans. Circuits Syst. {II}}, vol.~57, no.~2, pp. 69--74, 2010.

\bibitem{RTSC}
M.-H. Shen, J.-H. Tsai, and P.-C. Huang, ``Random swapping dynamic element matching technique for glitch energy minimization in current-steering {DAC},'' \emph{{IEEE} Trans. Circuits Syst. {II}}, vol.~57, no.~5, pp. 369--373, 2010.

\bibitem{913021}
M.~Rudberg, M.~Vesterbacka, N.~Andersson, and J.~Wikner, ``Glitch minimization and dynamic element matching in {D/A} converters,'' in \emph{Proc. 7th IEEE Int. Conf. Electron., Circuits and Syst. (ICECS)}, vol.~2, Dec. 2000, pp. 899--902.

\bibitem{9485113}
J.~Remple, A.~Panigada, and I.~Galton, ``An {ISI} scrambling technique for dynamic element matching current-steering {DACs},'' \emph{{IEEE} J. Solid-State Circuits}, vol.~57, no.~2, pp. 465--479, 2022.

\bibitem{4362086}
T.~Chen and G.~G.~E. Gielen, ``A 14-bit {200-MHz} current-steering {DAC} with switching-sequence post-adjustment calibration,'' \emph{{IEEE} J. Solid-State Circuits}, vol.~42, no.~11, pp. 2386--2394, 2007.

\bibitem{34100}
C.~Conroy, W.~Lane, and M.~Moran, ``Statistical design techniques for {D/A} converters,'' \emph{{IEEE} J. Solid-State Circuits}, vol.~24, no.~4, pp. 1118--1128, 1989.

\bibitem{1528680}
K.~Rafeeque and V.~Vasudevan, ``A new technique for on-chip error estimation and reconfiguration of current-steering digital-to-analog converters,'' \emph{{IEEE} Trans. Circuits Syst. {I}}, vol.~52, no.~11, pp. 2348--2357, 2005.

\bibitem{1692532}
Y.~Tang, H.~Hegt, and A.~van Roermund, ``{DDL}-based calibration techniques for timing errors in current-steering {DACs},'' in \emph{Proc. IEEE Int. Symp. Circuits Syst. (ISCAS)}, 2006, pp. 101--104.

\bibitem{5751593}
Y.~Tang, J.~Briaire, K.~Doris, R.~van Veldhoven, P.~C.~W. van Beek, H.~J.~A. Hegt, and A.~H.~M. van Roermund, ``A 14 bit {200 MS/s DAC} with {SFDR} $>$ 78 {dBc}, {IM3} $<$ -83 {dBc} and {NSD} $<$ -163 {dBm/Hz} across the whole {Nyquist} band enabled by dynamic-mismatch mapping,'' \emph{{IEEE} J. Solid-State Circuits}, vol.~46, no.~6, pp. 1371--1381, 2011.

\bibitem{AE_paper}
R.~Babaee, S.~Oveis~Gharan, and M.~Bouchard, ``Current-steering {DAC} architecture design for amplitude mismatch error minimization,'' in \emph{Proc. IEEE Int. Symp. Circuits Syst. (ISCAS)}, 2024.

\bibitem{optimization_book}
A.~J. Monticelli, R.~Romero, and E.~N. Asada, \emph{Modern Heuristic Optimization Techniques: Theory and Applications to Power Systems}.\hskip 1em plus 0.5em minus 0.4em\relax John Wiley \& Sons Ltd., 2008.

\bibitem{18626}
L.~Rabiner, ``A tutorial on hidden {Markov} models and selected applications in speech recognition,'' \emph{Proc. {IEEE}}, vol.~77, no.~2, pp. 257--286, 1989.

\end{thebibliography}
\bibliographystyle{IEEEtran}

\end{document}